\pdfoutput=1

\documentclass[11pt]{article}

\usepackage[final]{EACL2023}

\usepackage{times}
\usepackage{latexsym}

\usepackage[T1]{fontenc}

\usepackage[utf8]{inputenc}

\usepackage{microtype}

\usepackage{inconsolata}
\usepackage{graphicx}
\usepackage{tabularx}
\usepackage{multirow,multicol}
\usepackage{soul}
\usepackage{booktabs}
\usepackage{longtable}
\usepackage{url}
\usepackage{xcolor}

\title{HateModerate: Testing Hate Speech Detectors against Content Moderation Policies\\
\textcolor{red}{\it \small Warning: this paper discusses and contains content that can be offensive or upsetting.}}

\author{Jiangrui Zheng$^{\dagger}$, Xueqing Liu$^{\dagger}$\thanks{Corresponding author}, Guanqun Yang$^{\dagger}$, Mirazul Haque$^{\heartsuit}$, Xing Qian$^{\dagger}$, \\
\textbf{Ravishka Rathnasuriya$^{\heartsuit}$, Wei Yang$^{\heartsuit}$,  Girish Budhrani$^{\dagger}$}  \\
Stevens Institute of Technology$^{\dagger}$ \\
jzheng36,xliu127,xqian7,gyang16,gbudhran@stevens.edu\\ 
University of Texas at Dallas$^{\heartsuit}$ \\
mirazul.haque,ravishka.rathnasuriya,wei.yang@utdallas.edu }

\begin{document}
\maketitle
\begin{abstract}
To protect users from massive hateful content, existing works studied automated hate speech detection. Despite the existing efforts, one question remains: do automated hate speech detectors conform to social media content policies? A platform's content policies are a checklist of content moderated by the social media platform. Because content moderation rules are often uniquely defined, existing hate speech datasets cannot directly answer this question. 

This work seeks to answer this question by creating HateModerate, a dataset for testing the behaviors of automated content moderators against content policies. First, we engage 28 annotators and GPT in a six-step annotation process, resulting in a list of hateful and non-hateful test suites matching each of Facebook's 41 hate speech policies. Second, we test the performance of state-of-the-art hate speech detectors against HateModerate, revealing substantial failures these models have in their conformity to the policies. Third, using HateModerate, we augment the training data of a top-downloaded hate detector on HuggingFace. We observe significant improvement in the models' conformity to content policies while having comparable scores on the original test data. Our dataset and code can be found on \url{https://github.com/stevens-textmining/HateModerate}.
\end{abstract}

\section{Introduction}

\begin{figure}[h]
\centering
\includegraphics[width=8cm]{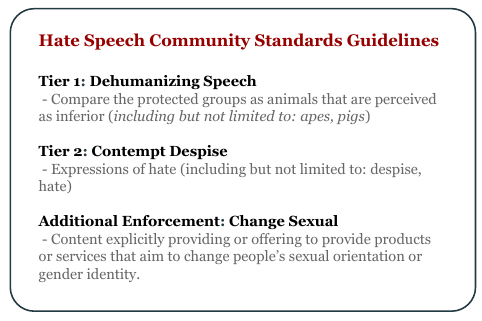}
\vspace{-5mm}
\caption{Examples of community standards guidelines for hate speech~\cite{fb_standard} \label{fig:intro}}
\vspace{-5mm}
\end{figure}

Social media platforms such as Facebook, Reddit, and Twitter/X have facilitated users to exchange information, but they also expose users to undesirable content, including hateful speech, misinformation, graphic violence, and pornography. To protect users from a massive amount of hateful content, existing work has been vigorously investigating new NLP approaches and providing new resources and open-source tools for studying hate speech detection~\cite{waseem2016hateful,davidson2017automated,vidgen2020learning,mathew2021hatexplain,hartvigsen2022toxigen,antypas2023robust}. 
Meanwhile, platforms also invested and achieved great success in building content moderation tools~\cite{facebook_report2,gpt4_as_content_moderator}, e.g., Facebook's automatic content moderator detected 95\% unwanted content before it is seen by a user~\cite{facebook_report2}.

Despite the existing work on hate speech, there remains an important question that is not well addressed: Do hate speech detectors' behaviors conform to platforms' content policies? Content policies are platform-specified rules on what content it moderates. For example, as of Nov 2022, Facebook specifies 41 community standards guidelines for moderating hate speech~\cite{fb_standard}; Figure~\ref{fig:intro} shows 3 examples of Facebook's guidelines. The content policies serve as a "contract" between users and the platform; without conforming to the policies, the decision on automated content moderators may be surprising to users, undermining the transparency and accountability of the moderation system. Such trustworthiness issues have led to incidents such as Reddit blackouts, which prevent users from accessing the contents normally~\cite{matias2016going}. Meanwhile, the answer to this question cannot be directly addressed using existing hate speech datasets. The reason is that many platforms have unique moderation rules, e.g., Facebook moderates advertisements on homosexual therapies. Our investigation shows that these custom rules are not well represented in existing hate speech datasets, causing an underestimation of the models' failures in conforming to these rules.

To assess the conformity of automated content moderators to content policies, this paper proposes a dataset called HateModerate, which consists of 7.6k hateful and non-hateful examples for the 41 community standards guidelines on Facebook. Among the published moderation rules from existing work~\cite{banko2020unified,fb_standard,rottger2020hatecheck}, we opt for Facebook's community standards guidelines for hate speech~\cite{fb_standard} as previous work shows it is the most comprehensive among all platforms~\cite{Jiang2020Characterizing} and it has good clarity. 



\begin{figure*}[h]
\vspace{-13mm}
\centering
\includegraphics[width=\linewidth]{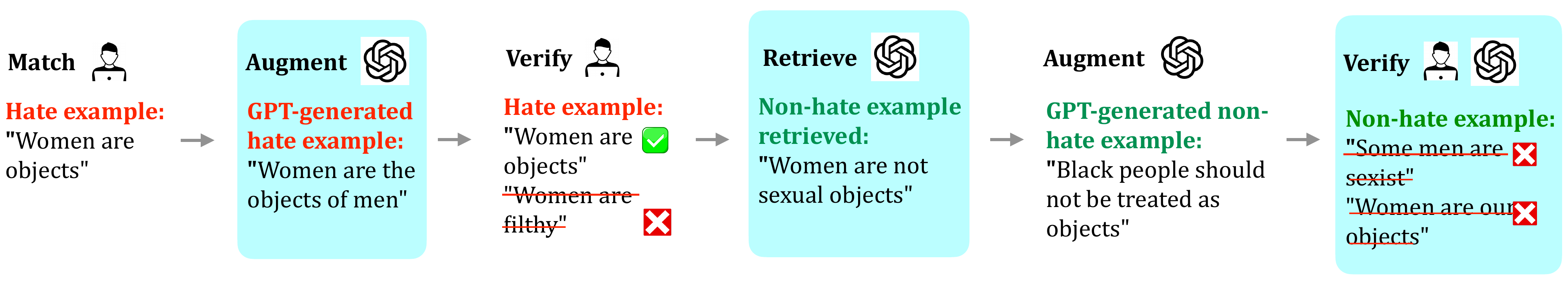}
\caption{The workflow of data collection for \textsf{Guideline 10 (Tier 1, Certain objects)}. \label{fig:workflow}}
\vspace{-5mm}
\end{figure*}

HateModerate is constructed using the six-step process illustrated in Figure~\ref{fig:workflow}. First, we recruit a group of 28 graduate students as the annotators. A part of these students manually search for hateful examples from existing datasets matching each policy. Second, since some guidelines contain too few matched examples, we augment these guidelines by generating hateful examples with the GPT engine. Third, to ensure that the searched and generated examples indeed match the criteria, 16 additional annotators manually verify each hateful example. Fourth, after the hateful examples are collected, for each guideline, we retrieve difficult non-hateful examples from existing datasets that closely resemble the hateful examples to help detecting the model failures. Fifth, similarly, we augment guidelines with GPT-generated non-hateful examples. Sixth, 4 additional annotators manually verify each non-hateful examples. The average agreement rate (Krippendorf's alpha) on the match/unmatch of hateful and non-hateful examples are 0.521 and 0.809.


After constructing HateModerate, we examine state-of-the-art hate speech detectors against each policy using the dataset. More specifically, we examine the following models: Google's Perspective API~\cite{perspective_api}, OpenAI's Moderation API~\cite{openai_moderation}, Facebook's RoBERTa model~\cite{fb_moderation_api_r1} and Cardiff NLP's RoBERTa model~\cite{antypas2023robust}. We make the following observations. First, all models prioritize more severe policies (e.g., violence) compared to less severe policies (e.g., stereotyping); second, the OpenAI model conforms the best to the content policies; third, besides OpenAI, models generally have high failure rates for non-hateful examples. After observing the model failures, we further seek answers on how to improve the models' conformity to policies. By adding HateModerate to the training dataset of a top-downloaded model on HuggingFace, the model's performance on HateModerate and HateCheck~\cite{rottger2020hatecheck} is significantly improved while the performance on the original test set remains comparable. We also conduct a fine-tuning experiment comparing HateModerate and DynaHate, which demonstrates HateModerate's out-performance over existing datasets.
These results highlight the importance of our dataset in improving the model conformity to content policies. 


\section{Background and Related Work}
\label{sec:relwork}

\subsection{Hate Speech Detection}


\noindent \textbf{Construction of Hate Speech Datasets}. Automatically detecting hateful speech online is a challenging problem in natural language processing. In recent years, hate speech detection benefits from the advancement of machine learning and NLP techniques~\cite{he2023youonly,gpt4_as_content_moderator}; nevertheless, previous work argues that the datasets play a more important role than the model architecture in hate detection~\cite{grondahl2018all}. Existing work has contributed to many public datasets for hate speech detection~\cite{waseem2016hateful,davidson2017automated,vidgen2020learning,mathew2021hatexplain,hartvigsen2022toxigen}. Since hate speech constitutes approximately 1\% of all online speech~\cite{sachdeva2022measuring}, previous work leverage different sampling techniques to improve the efficiency of labeling. For example, by using pre-defined keywords and Twitter hashtags~\cite{davidson2017automated,he2021racism,waseem2016hateful,golbeck2017large}. However, hard filtering based on keywords may lead to low coverage issues~\cite{sachdeva2022measuring}. Alternatively, previous work employed information retrieval~\cite{rahman2021information} and classification to create a soft filter~\cite{sachdeva2022measuring}. Our work does not have the class imbalance problem as we reuse the existing hate speech datasets. We further improve the coverage of the dataset with GPT-generated examples. 


\noindent \textbf{The Taxonomy for Hate Speech Detection}. A taxonomy defines what content is considered hateful. A taxonomy with detailed guidelines can help non-expert annotators better understand the labeling goal. The guidelines contain a checklist of descriptions of the hateful and non-hateful content~\cite{waseem2016hateful,sachdeva2022measuring,elsherief2021latent}; some previous work further provides codebooks containing more detailed instructions on what is not considered as hateful for each guideline~\cite{golbeck2017large,vidgen2020learning}. 
Banko et al.~\cite{banko2020unified} introduce a unified taxonomy of harmful content, including \emph{sexual aggression, doxxing, misinformation} and \emph{hate speech}. Our annotators are provided with Facebook's 41 community standards guidelines. These guidelines contain fine-grained categories (e.g., subcategories of dehumanization) of hate speech as well as new categories that are not well covered in existing datasets (e.g., advertisements of homosexual therapies).

\subsection{Policies for Content Moderation}

\label{sec:analysis}

\noindent \textbf{Regulations of Governments/Councils}. Online content moderation is subject to policies and regulations of the governments~\cite{section230,dsa}. \citet{zufall-etal-2022-legal} constructs a "punishable" hate speech dataset in German based on the German Criminal Code and a legal decision framework. \citet{chiril-etal-2021-nice-wife} study gender bias based on the definition by the French High Council on Gender Equality.  

\noindent \textbf{Social Media Content Policies}. Although platforms have the right to decide what content to moderate~\cite{section230}, users show concerns over the consistency and transparency of the moderation decisions~\cite{matias2016going}. To improve the transparency of moderation, many major platforms released their content policies~\cite{fb_standard,twitter_policies,ig_standard,pinterest_standard,reddit_standard}, which serve as a "contract" between the user and the moderation system. 
The policies are based on what value is preserved by the platform, which vary across platforms, e.g., Gab allows more elitism speeches than Twitter~\cite{Zhou_Dredze_Broniatowski_Adler_2019}. \citet{Jiang2020Characterizing} conduct a comparative study of the existing community standards guidelines across platforms; their study suggests that Facebook's guidelines are the most comprehensive ones above all.


\noindent \textbf{Facebook Community Standards}. As of Nov 2022, Facebook provides a list of 41 community standards guidelines for hate speech moderation~\cite{fb_standard}. Figure~\ref{fig:intro} shows three examples of Facebook's hate speech guidelines, and Table~\ref{tab:fb_standards} shows the complete list. Facebook's guidelines are organized into four tiers based on the content severity~\cite{fb_standard}: Tier 1 includes the most offensive content,
e.g., dehumanization and violence towards protected groups; Tier 2, Tier
3, and Tier 4 (the additional enforcement) are less severe,
e.g., stereotyping and contempt towards protected groups. 
In this work, we leverage Facebook's community standards guidelines for constructing our dataset.

\subsection{Behavioral/Capability Tests of NLP Models}


HateModerate provides fine-grained failure rate estimation for each content policy. To this end, it can be seen as a dataset for \emph{capability tests}~\cite{ribeiro2020beyond,rottger2020hatecheck,yang2022testaug}. The traditional held-out tests may overestimate the model performance when the model has bias~\cite{poliak2018hypothesis}. To alleviate this issue, \citet{ribeiro2020beyond} propose to construct a checklist of out-of-domain test suites for each capability the model should have. In particular, HateCheck~\cite{rottger2020hatecheck} provides a list of 29 test suites for hateful and non-hateful capabilities, e.g., "\emph{We are a group of [PROTECTED GROUP].}" is a non-hateful suite. However, most of the test suites of HateCheck focus on defining hate speeches with \emph{syntactic} structures, and HateCheck's rules suffer from a low coverage of the hate speech categories (Section 4.3 of \citet{rottger2020hatecheck}). On the other hand, the test suites of HateModerate focus on semantic categories specified by the guidelines; it also improves the coverage of hateful content compared to HateCheck.

\section{Constructing the HateModerate Dataset}
\label{sec:step1}

In this section, we describe the steps for the construction of HateModerate.

\noindent \textbf{Annotators Recruitment}. HateModerate is annotated by 28 graduate students in Computer Science.\footnote{We opt for students labeling rather than Amazon Mechanical Turk labeling since the quality of students' labeling is more manageable, we notice some existing work on hate speech dataset collection also used students labeling~\cite{fanton-etal-2021-human}. } The annotators are recruited from PhD and Master students at a research lab and students taking a graduate-level NLP course. The annotation process is overseen by two experts in online hate. All participants are compensated with a \$20 Amazon e-gift card. The annotator names are anonymized in the dataset. We obtained the annotators' consent, and it was explained to the annotators how the data would be used. More details about the annotator recruitment can be seen in Section~\ref{sec:ethics}.

\noindent \textbf{Data Sources}. Most of Facebook's community standards guidelines are on general hateful content, e.g., dehumanization. Therefore, existing datasets should already contain examples matching a significant number of guidelines. We thus first try to search for and reuse examples and their hateful/non-hateful labels from existing datasets. By doing so, we reduce the requirement on annotator expertise and avoid introducing additional labeling errors; notably, it is challenging for non-expert annotators to reach a high agreement rate on hateful/non-hateful labels~\cite{mathew2021hatexplain}. We first instruct the annotators to search in the following datasets: DynaHate~\cite{vidgen2020learning}, Toxic Spans~\cite{pavlopoulos-etal-2021-semeval}, Hate Offensive~\cite{davidson2017automated}, and HateCheck~\cite{rottger2020hatecheck}. Later the annotators extended the list to include Twitter Hate Speech~\cite{twitter_hate}, Ethos~\cite{mollas2020ethos}, FRENK~\cite{ljubevsic2019frenk}, and COVID Hate and Counter Speech~\cite{he2021racism}. The hateful/non-hateful labels are available in all datasets. 

\subsection{Collecting Hateful Examples}
\label{sec:collection}
 
\noindent \textbf{Manually Searching Matching Hateful Examples}. For the first step, we collect the hateful examples matching each guideline. We assign each of Facebook's 41 policies to one of 7 annotators. Annotators are instructed to search for a minimum of 200 hateful examples. Synonyms and regular expressions are employed to enhance the search efficiency. For example, for \textsf{Guideline 0 (Tier 1, Dehumanize Filth}), the annotator uses the regular expression "\emph{.*(filth$\vert$dirt).*}". Multiple annotators also report that they seek help from Google, ChatGPT, and other team members to correctly understand the policy. For example, for \textsf{Guideline 13 (Tier 1, Hatecrime Victim)}, the assigned annotator first struggles to find enough examples because the exact word "\emph{hate crime}" rarely appears in any example, after a discussion, the team members suggest him to search for concrete hate crime keywords including \emph{lynching} and \emph{holocaust}. The annotator is able to add more examples as a result. 
%

\noindent \textbf{Data Augmentation for Hateful Examples}. After the initial manual search, we find that for some guidelines, the existing datasets do not contain a sufficient number of matching examples, e.g., our annotators only find 3 matching examples for \textsf{Guideline 38 (Tier 4, Change Sexual)}. To improve the balance of these guidelines, we further ask the initial annotator to either manually create or use the GPT-3 engine (\textsf{text-davinci-003}) to generate synthetic hateful examples. The prompt we used for the generation is: "\emph{Based on hate speech guideline [GUIDELINE], write a hate speech sentence about [PROTECTED GROUP]}".


\noindent\textbf{Manually Verifying Matched Hateful Examples}. In the first step, the hateful examples in each guideline are matched by only one annotator. We need further verification to reduce the errors in the initial annotators' matching. To this end, we involve a total of 16 additional annotators to label whether an initially found example matches (1) or does not match (0) the guideline. We remove an example if at least one annotator labels it as 0. Each example is labeled by 1 to 7 annotators. Following existing work on using demonstrated examples to improve the quality of crowdsourced annotations~\cite{gupta2022ezcoref}, we provide a few false match examples for each policy, e.g., for \textsf{Guideline 29 (Tier 2, Curse Sexual)}, "\emph{They make me so pissed off these immigs!!}" is demonstrated as a false match. The average agreement rate (Krippendorf's alpha) between the annotators is 0.521 (Appendix~\ref{sec:agreement}). 









\begin{figure}[t]
\centering
\includegraphics[width=1\linewidth]{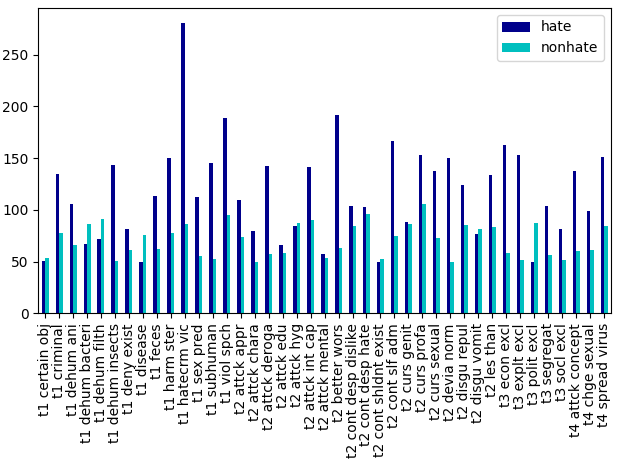}
\vspace{-8mm}
\caption{The statistics of examples in each policy in our dataset\label{fig:stats}}
\vspace{-5mm}
\end{figure}

As mentioned above, some guidelines contain few matching examples from existing datasets, we repeatedly perform verification/removal and augmentation until each guideline contains at least 50 valid matching hateful examples.

\subsection{Collecting Non-Hateful Examples}\label{sec:collec_nonhate}

\noindent \textbf{Retrieving Difficult Non-Hateful Examples}\label{sec:hard_negative}. Since testing with only hateful examples will result in bias (e.g., one model has a low failure rate simply because it sets a low threshold for hate), we further add non-hateful examples to HateModerate. To improve the detection of model failures, for each policy, we opt for retrieving more difficult non-hateful examples that are most similar to the hateful examples from the previous stage. The corpus we retrieve from are the non-hateful examples in DynaHate~\cite{vidgen2020learning}, since DynaHate contains a large number of manually created adversarial non-hateful examples that look similar to hateful examples. The retrieval algorithm follows the state-of-the-art dense retrieval paradigm~\cite{karpukhin2020dense}. We employ OpenAI's Embedding API~\cite{OpenAI_embeddings} with the \texttt{text-embedding-ada-002} model to obtain the vectors. For each policy, we rank every non-hateful example in DynaHate by its average cosine similarity with the existing hateful examples and keep the top-100 non-hateful examples in this step.  

\noindent \textbf{Data Augmentation for Non-Hateful Examples}. Similar to hateful examples, DynaHate does not contain enough non-hateful examples matching some guidelines. We thus also perform data augmentation for non-hateful examples. First, we use GPT-3 (\textsf{text-davinci-003}) to generate non-hateful examples using the following prompt: "\emph{Based on hate speech guideline [GUIDELINE], write a sentence about [PROTECTED GROUP] with [NON-HATE TYPE]. Examples: [EXAMPLES].}". In particular, the data augmentation re-balances the non-hateful type (explained below), i.e., supporting, counter-hate, neutral, and offensive speech against non-protected groups. For offensive speech against non-protected groups, it is difficult for GPT-3 to generate matching examples, so we manually create the non-hateful examples. 

\noindent \textbf{Verifying Non-Hateful Examples}. Similarly, the retrieved and augmented non-hateful examples may not closely match the guideline. For example, for \textsf{Guideline 10 (Tier 1, Certain objects)} on dehumanizing speech as objects, one top-retrieved non-hateful example is: "\emph{Some men are sexist}" which is unrelated to the guideline. For each example, we further involve four annotators to provide labels on whether one example is related to the guideline (1) or not (0). Each example receives 2 labels. We remove an example if at least one annotator labels it as 0. The average agreement rate (Krippendorf's alpha) between the annotators is 0.809 (Appendix~\ref{sec:agreement}). 

We further perform the following classification step for the non-hateful examples. For each non-hateful example, we employ GPT-4 and 1 annotator's verification to classify it into five classes: supporting, counter-hate, neutral, offensive speech against non-protected groups, and hateful speech with the wrong label.\footnote{The prompt we used for GPT-4 classification is: "\emph{Classify the sentence of Question into categories 1-5, number only + [GUIDELINE]+[EXAMPLES]}".} The first three classes are based on the definition of non-hateful speeches in previous work~\cite{sachdeva2022measuring}, and we identify the 4th class during labeling. The full descriptions of the five classes can be found in Appendix~\ref{sec:class_nonhate}. This classification step allows us to remove the hateful examples wrongly labeled as non-hateful (about 3.6\%) and to re-balance the four non-hateful types in the data augmentation. 



\subsection{The Agreement Rates between Annotators}
\label{sec:agreement}

Table~\ref{tab:agreement} includes detailed agreement rates on verifying whether an example matches or does not match a guideline. We report  Krippendorf's $\alpha$ which is often used in previous work on crowd-sourcing~\cite{mathew2021hatexplain,vidgen2020learning} and the ratio of agreement. 

\begin{table}[h]
\caption{The inter-annotators agreement rates and Krippendorff's $\alpha$ in the HateModerate validation process.}
\label{tab:agreement}
\centering
\small
\begin{tabular}{p{3.5cm}p{1.5cm}p{1.5cm}}

HateModerate & \textbf{Hate} & \textbf{Non-Hate} \\
\hline
Ratio of Agreement & 89.64\% & 91.15\% \\
Krippendorff's $\alpha$ (Nominal) & 0.521 & 0.808\\
Krippendorff's $\alpha$ (Interval) & 0.521 & 0.809 \\
\hline
\end{tabular}
\end{table}

\subsection{Dataset Statistics}

In our final HateModerate dataset, we compile 7,704 examples: 4,796 hateful (4,535 unique ones) and 2,908 non-hateful (2,264 unique ones). Some instances are duplicated because a single sentence can fall under multiple guidelines simultaneously. The majority of examples come from DynaHate (5,174), followed by GPT (1,385), HateCheck (457), manual (270), Toxic Span (102), COVID hate (152), Hate Offensive (92), Ethos (12), Twitter Hate (33), Toxigen (8) and FRENK (19).



Figure~\ref{fig:stats} shows the statistics of HateModerate by policy. Among the 41 policies, the most frequent policy contains 367 examples whereas the least frequent policy contains 103 examples, all policies contain 100 to 250 examples, and the majority policies contain more than 150 examples. We demonstrate how diverse the hate speech and non-hate speech samples are in terms of semantics, vocabulary, and length statistics for each sample, as shown in Table~\ref{tab:analys_data}.

\begin{table}[h]
\caption{The analysis of vocabulary size, average number of tokens, and median number of tokens of the HateModerate dataset.}
\label{tab:analys_data}
\centering
\small

\begin{tabular}{p{2cm}p{2cm}p{1cm}p{1cm}}
\hline
HateModerate&  Vocab Size&  Avg. & Median \\
\hline
All&  11,775&  20.98& 14\\
Hate&  9,869&  22.57& 15\\
Nonhate&  5,518&  18.35& 12\\
\hline
\end{tabular}
\end{table}

\begin{figure*}[t]
\centering
\vspace{-5mm}
\includegraphics[width=\linewidth]{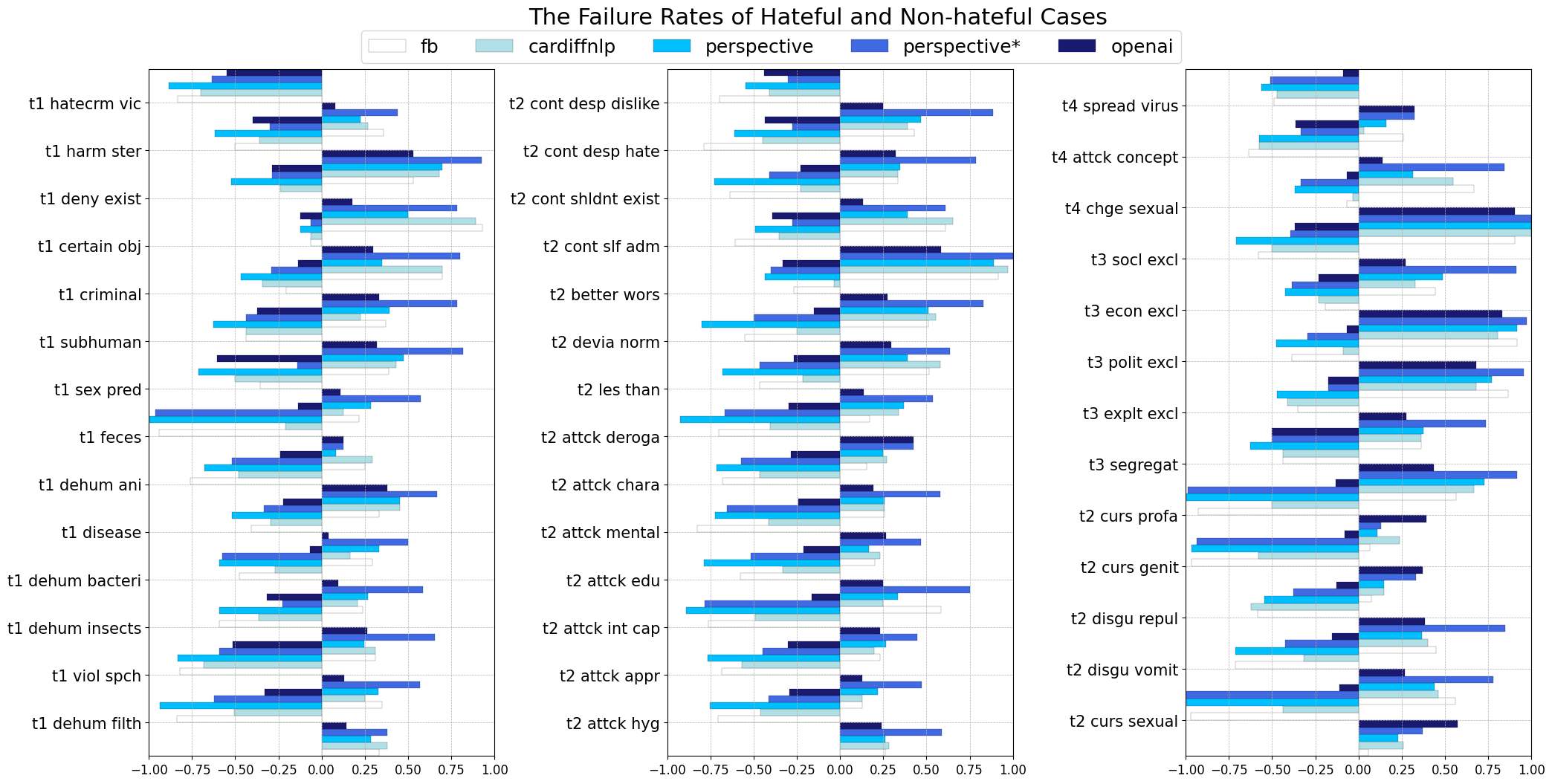}
\caption{We detect the failure rates for both hateful and non-hateful examples across each of the 41 policies in Facebook's community standards guidelines~\cite{fb_standard}. Perspective's threshold is 0.5; Perspective*'s threshold is 0.7. For each policy, the bars facing right show the failure rates of hateful examples; the bars facing left show the failure rates of non-hateful examples.  \label{fig:fail_rate_all}}
\end{figure*}

\begin{table*}[h]
\caption{The average failure rates of the hateful and non-hateful examples for different tiers of policies, and the average toxicity scores. F: Facebook model, C: Cardiff NLP, P: Perspective with threshold 0.5, P*: Perspective with threshold 0.7, O: OpenAI's API.  \label{tab:sum} }
\vspace{-3mm}
\begin{center}
\small
\begin{tabular}{c|p{.27cm}|p{.27cm}|p{.27cm}|p{.27cm}|p{.27cm}|p{.27cm}|p{.27cm}|p{.27cm}|p{.27cm}|p{.27cm}|p{.27cm}|p{.27cm}|p{.27cm}|p{.27cm}|p{.27cm}|p{.27cm}|p{.27cm}|p{.27cm}|p{.27cm}|p{.27cm}|p{.27cm}|p{.27cm}}
\hline
\multirow{3}{*}{\textbf{T}} & \multicolumn{12}{c|}{Failure Rate} & \multicolumn{10}{c}{Average Toxicity Score} \\ \cline{2-23}
& \multicolumn{6}{c|}{Hate} & \multicolumn{6}{c|}{NonHate} & \multicolumn{5}{c|}{Hate} & \multicolumn{5}{c}{NonHate} \\ \cline{2-23}
& avg & F & C & P & P* & O & avg & F & C & P & P* & O & avg & F & C & P & O & avg & F & C & P & O \\ \hline
\textbf{1} & .34& .40& .38& .35&.62& \textbf{.22}& .47& .52& .39& .65& .43&  \textbf{.31}& .64& .62& .65& .54& \textbf{.74}& .44& .58& .43& .47& \textbf{.27}\\ \hline
\textbf{2} & .34& .34& .37& .34& .60& \textbf{.30}& .52& .69& .40 & .74& .55& \textbf{.24}& .62& .64& .62& .55& \textbf{.66}& .47& .71& .43& .55& \textbf{.21}\\  \hline
\textbf{3} & .59& .63& .57& .66&.90& \textbf{.50}& .38& .39& .33& .54& .35&  \textbf{.27}& .48& .45& .50& .43& \textbf{.55}& .33& .42& .31& .38& \textbf{.19}\\  \hline
\textbf{4} & .52& .61& .53& .49&.72& \textbf{.46}& .36& .40& .36& .50& .39& \textbf{.17}& .52& .41& .50& .50& \textbf{.68}& .35& .44& .38& .44& \textbf{.14}\\  \hline
\end{tabular}
\end{center}
\vspace{-4mm}
\end{table*}

\section{Testing Hate Speech Detectors' Conformity with Content Policies}
\label{sec:step2}

In this section, we employ HateModerate as our evaluation benchmark to assess how hate speech detectors conform to content policies. We seek answers to the following research questions: 

\noindent \textbf{RQ1: How do popular and commonly used hate speech detectors conform to Facebook’s content policies?} 

\noindent\textbf{RQ2: What policies do hate speech models conform to the least?} 

After our initial evaluation, we observe that state-of-the-art models all had different degrees of failures conforming to the content policies. To understand if such failures can be alleviated, we further try fine-tuning existing models with HateModerate. We ask the following research questions:

\noindent\textbf{RQ3: Does adding HateModerate to the training data help improve a model's conformity to content policies?} 

\noindent\textbf{RQ4: How Novel is HateModerate Compared to Existing Datasets?}

\subsection{Experiment Setup}

\noindent \textbf{Hate Speech Models Evaluated}. To answer RQ1-RQ3, we evaluate state-of-the-art models from both industry API endpoints and open-source hate speech detection models. For industry APIs, we choose Google's Perspective API~\cite{perspective_api} and OpenAI's Moderation API~\cite{openai_moderation,markov2022holistic}, which are frequently used in downstream detection tasks~\cite{alpaca,perspective_api_use}; for open-source models, we choose Cardiff NLP's fine-tuned RoBERTa model~\cite{antypas2023robust} and Facebook's Fine-Tuned RoBERTa model~\cite{fb_moderation_api_r1} which rank top-2 and top-1 among the most downloaded hate models on HuggingFace. The full details of the models can be found in Appendix~\ref{sec:models}. 

\noindent \textbf{Train/Test Split and Avoiding Data Contamination}. To answer RQ3, we reserve 50\% of HateModerate for fine-tuning (cf. Section~\ref{sec:augmentation}) by random sampling and use the other half for testing. One issue with evaluating the above models is that their training data may overlap with HateModerate testing data, causing unfair comparisons between models. To minimize the impact of the potential data contamination, for the testing fold, we keep only newly created datasets that are not in the training data of any models. The full details of the excluded data can be found in Appendix~\ref{sec:remove_original}.



\noindent \textbf{Evaluation Metric}. Following previous work on capability testing~\cite{rottger2020hatecheck,ribeiro2020beyond}, we report the average failure rate of the hateful and non-hateful examples in each policy. If the hateful failure rate is high, it indicates the model cannot effectively detect this category of hate speech; if the non-hateful failure rate is high, it indicates the model cannot effectively recognize non-hateful speeches for that category.

\subsection{Evaluating Model Failures using HateModerate}
\label{sec:testing_results}

\noindent In this section, we seek answers to RQ1 and RQ2. We report the failure rates of each policy in Figure~\ref{fig:fail_rate_all}. In addition, we report the average failure rate and the average toxicity score of each model in each tier in Table~\ref{tab:sum}. From Figure~\ref{fig:fail_rate_all} and Table~\ref{tab:sum} we can make the following observations. 

\subsubsection{Comparison of Model Failures of different Policies}

\noindent From Figure~\ref{fig:fail_rate_all} and Table~\ref{tab:sum}, we can observe a contrastive trend between higher tier policies (i.e., more severe hate speech) and lower tier policies. In general, for higher tier policies, the failure rates for hateful examples are lower; whereas the failure rates for non-hateful examples are higher. These results imply that, the less severe speeches specified by lower tier policies are not prioritized by both the industry APIs and the open-source models.

\noindent \textbf{Case Studies of Individual Policies with High Failure Rates}. To confirm the validity of our observation, we further conduct case studies on individual policies with the highest failure rates of hateful examples. We identify that most models fail to detect hate speech specified by the following policies: \emph{Certain Objects, change sexual, economic exclusion} and \emph{self admission to intolerance of protected characteristics}. For example, most models fail to detect "\emph{MAKE WOMEN PROPERTY AGAIN!!!!}". We describe these failures in details in Appendix~\ref{sec:casestudy}.

\subsubsection{Comparison of Different Models' Failures} 

By comparing the failure rates over all examples (hateful and nonhateful, all tiers) in Table~\ref{tab:sum}, we observe that OpenAI's model exhibits the lowest average failure rate (avg: 0.29, std: 0.17), followed by Perspective (avg: 0.38, std: 0.19). CardiffNLP  (avg: 0.40, std: 0.22) and  Facebook's RoBERTa (avg: 0.40, std: 0.23) perform less well.

Besides OpenAI, most of the models exhibit high failure rates in non-hateful examples. Perspective with 0.5 threshold performs the worst in non-hateful examples. We further report the failure rate of Perspective with 0.7 threshold in Table~\ref{tab:sum}. We can observe a trade-off between good failure rates in the hateful and non-hateful examples of the two thresholds. 


\noindent \textbf{Bias in Toxicity Scoring}. In Table~\ref{tab:sum}, we report the average toxicity scores of each model for different tiers of policies, i.e., the probability for the model to predict the hateful class. We can see that while different models have similar toxicity scores for the hateful examples, the scores for non-hateful examples are different. Essentially, Perspective and Facebook's RoBERTa tends to assign higher toxicity for both hateful and non-hateful examples.

\noindent \textbf{Finding Summary of RQ1 and RQ2}. \textcircled{1} All models prioritize more severe policies over less severe policies; \textcircled{2} The OpenAI model has the best performance overall, Perspective generally scores sentences with higher toxicity scores, thus a threshold higher than 0.5 is desirable; 
\textcircled{3} The models are generally bad at detecting difficult non-hateful examples except for OpenAI (a more detailed analysis can be found in Appendix~\ref{sec:sub_nonhate}).

\begin{table}[h]
\centering
\caption{The failure rates of fine-tuning with the CardiffNLP data before and after adding HateModerate. Significant results are denoted with $\dagger$.\label{tab:fine_tune_roberta}}
\label{table:evaluation}
\small
\renewcommand{\arraystretch}{1.0}
\setlength\tabcolsep{3.5pt} 
\begin{tabular}{p{1.5cm}cccr}

\toprule
\textbf{FailureRate} & \multicolumn{3}{c}{\textbf{Fine-tuned RoBERTa on}} \\
\cmidrule(l){2-4}
& \textbf{CardiffNLP} & \textbf{+ HM} & \textbf{+ HM*} & \textbf{OpenAI}\\
\midrule
\multicolumn{3}{l}{\textit{\textbf{HateCheck~\cite{rottger2020hatecheck}}}} \\
Hate & .442& .185$^{\dagger}$ & .297$^{\dagger}$ & \textbf{.008}\\
Non-hate & .205& .229$^{\dagger}$ & .205 & \textbf{.016}\\
Overall & .365& .199$^{\dagger}$ & .235$^{\dagger}$ & \textbf{.011}\\
\midrule
\multicolumn{3}{l}{\textit{\textbf{HateModerate Test}}} \\
Hate & .454& \textbf{.222}$^{\dagger}$ & .281$^{\dagger}$ & .369\\
Non-hate & .409& .338$^{\dagger}$ & \textbf{.301}$^{\dagger}$& .351\\
Overall & .423& \textbf{.275}$^{\dagger}$ & .295$^{\dagger}$ & .365\\
\midrule
\multicolumn{3}{l}{\textbf{CardiffNLP Test Sets:}} \\
\multicolumn{3}{l}{\textit{\textbf{hatEval~\cite{basile-etal-2019-semeval}}}} \\

Hate & .084& .075 & \textbf{.061}$^{\dagger}$ & .754\\
Non-hate & \textbf{.776} & .781 & .780 & .080\\
Overall & .485 & .485 & .478$^{\dagger}$ & \textbf{.363}\\
\multicolumn{3}{l}{\textit{\textbf{HTPO~\cite{grimminger-klinger-2021-hate}}}} \\
Hate & .526 & .661$^{\dagger}$ & \textbf{.525}& .949\\
Non-hate & .043 & .037 & .041& \textbf{.006}\\
Overall & .090 & .090$^{\dagger}$ & \textbf{.089} & .098\\
\multicolumn{3}{l}{\textit{\textbf{HateXplain~\cite{mathew2021hatexplain}}}} \\
Hate & \textbf{.157} & .159 & .168& .351\\
Non-hate & .315 & .262$^{\dagger}$ & .266$^{\dagger}$ & \textbf{.223}\\
Overall & .221 & \textbf{.201} $^{\dagger}$& .208$^{\dagger}$ & .299\\
\bottomrule
\end{tabular}
\vspace{-4mm}
\end{table}

\subsection{Mitigating Model Failures with Fine-Tuning HateModerate}
\label{sec:augmentation}

In this section, we seek the answer to RQ3. We do so by comparing the failure rates of the following models in Table~\ref{table:evaluation}: \textcircled{1} \textbf{CardiffNLP}: RoBERTa-base fine-tuned using all the available training data for the CardiffNLP model~\cite{antypas2023robust};\footnote{We are only able to access 9 out of the 13 training datasets of the CardiffNLP model. The full details of 9 datasets can be found in Appendix~\ref{sec:card_datasets}.} \textcircled{2} \textbf{+HM}: RoBERTa-base fine-tuned using CardiffNLP's training data + HateModerate's reserved training data; \textcircled{3} \textbf{+HM$^*$}: same as \textbf{+HM} but downsample the hateful examples so the hateful and non-hateful examples are balanced; \textcircled{4} \textbf{OpenAI}: The failure rate of the OpenAI API.
For the 9 training datasets of the CardiffNLP model, we use the same train/test split as the original datasets.\footnote{Among all 9 datasets, the train/test split is available in only 3 datasets, which we use as the test sets in Table~\ref{tab:fine_tune_roberta}. We use all remaining data for the train.}  The hyperparameters and more details of fine-tuning can be found in Appendix~\ref{sec:finetune_process}.

\noindent \textbf{Results of Fine-Tuning}. In Table~\ref{tab:fine_tune_roberta}, we compare the failure rates on the following test collections: \textcircled{1} The testing fold of HateModerate; \textcircled{2} The 3 testing datasets of CardiffNLP; \textcircled{3} HateCheck~\cite{rottger2020hatecheck}, a dataset for independent out-of-domain capability tests of hate speech. 
We conduct the paired t-test between \textbf{+HM} vs \textbf{CardiffNLP} and \textbf{+HM$^*$} vs \textbf{CardiffNLP}. In the \textbf{+HM} and \textbf{+HM$^*$} columns, we denote the significant results (p-value < 0.05) using $\dagger$. The details of the t-test results can be found in Table~\ref{tab:ttest} of Appendix~\ref{sec:ttest}. 
Table~\ref{tab:fine_tune_roberta} reveals that adding HateModerate to the fine-tuning set significantly reduces the failure rates on HateModerate and HateCheck, while the failure rates on the CardiffNLP's test sets are comparable. While adding \textbf{+HM} sometimes make the non-hate failure rate even worse than \textbf{CardiffNLP}, re-balancing the hateful and non-hateful examples can alleviate this problem. Furthermore, while OpenAI performs the best in Table~\ref{tab:sum} and Figure~\ref{fig:fail_rate_all}, in Table~\ref{table:evaluation} it has higher failure rates than \textbf{+HM} and \textbf{+HM$^*$} on the HateModerate test. This comparison with the strong OpenAI model further confirms the significance of our dataset. 

Previous work has shown that fine-tuning hate speech models can lead to bias, e.g., the two-word sentence "\emph{black women.}" is predicted as hateful~\cite{markov2022holistic,reddit_api,zhou-etal-2021-challenges,davidson-etal-2019-racial}. We perform an analysis of measuring such bias in our fine-tuned model in Appendix~\ref{sec:bias_check}.



\noindent \textbf{Finding Summary of RQ3}. 
We find that by fine-tuning hate speech detection models with HateModerate, we can effectively reduce the models' non-conformity to content policies.

\vspace{-0.1in}

\subsection{Measuring the Bias of the Fine-Tuned Models}
\label{sec:bias_check}

Prior work shows that fine-tuning hate speech detectors can lead to bias against certain protected groups, e.g., the two-word sentence "\emph{black women.}" is predicted as hateful. Previous work thus measures such bias when fine-tuning a hate detection model~\cite{markov2022holistic,reddit_api}. For example, Markov et al.~\cite{markov2022holistic}
 use 69k curated synthetic examples like "\emph{[subject] is selfish/foolish/narrow-minded.}" to measure and mitigate the bias. 

Following the previous work, we also measure the bias in this paper. We test the 3 fine-tuned models in Table 2 (i.e., \textbf{CardiffNLP}, \textbf{+HM}, \textbf{+HM$^*$}) with 21 synthetic non-hateful examples and 13 hateful examples such as "\emph{[PROTECTED GROUP] ARE [POS/NEG ADJ].}". Surprisingly, almost 100\% of the non-hateful predictions are correct while 15\% of the hateful predictions are incorrect. In Table~\ref{table:simple_test}, we report the only test cases in which we find the hate probability of the fine-tuned model is abnormal. 

Besides the simple examples in Table~\ref{table:simple_test}, we further measure the bias using more realistic examples from HateCheck~\cite{rottger2020hatecheck}. HateCheck contains 18 hateful and 11 non-hateful suites of test cases on 7 protected groups. We find the 3 fine-tuned models generally have low failure rates on the non-hateful examples of HateCheck. In Table~\ref{table:hatecheck}, we report all test suites in HateCheck whose failure rates are higher than 50\%, including two test suites about women. To study whether adding HateModerate increases the bias compared to the original model, we further perform the paired t-test between \textbf{CardiffNLP} vs \textbf{+HM}'s predictions on HateCheck non-hateful examples (p-value: 0.80), and \textbf{CardiffNLP} vs \textbf{+HM$^*$} (p-value: 0.83). Since the p-values are not significant, we can reject the null hypothesis that HateModerate introduces more bias to the model.  

 \begin{table}[h]
\caption{Measuring the bias: all test suites in HateCheck whose failure rates are higher than 50\% \label{table:hatecheck} }
\begin{center}
\vspace{-5mm}
  \label{table:hatecheck}
  \begin{tabular}{p{3cm}p{1cm}p{0.6cm}p{0.6cm}p{0.6cm}}
    \toprule
    Test Suite  &Group& \textbf{Card} & \textbf{+HM} & \textbf{+HM$^*$} \\
    \hline
 \textbf{F8}: Non-hateful homonyms of slurs &Women& .80& .80&.70\\
  \textbf{F9}: Reclaimed slurs&Women& .47& .67&.60\\
    \textbf{F23}: Abuse targeted at individuals
(not as member of a prot. group) & None & .45& .46&.52\\
  \textbf{F24}: Abuse targeted at nonprotected groups (e.g. professions) &Non-Protected Group& .58& .52&.58\\
  \hline
  \end{tabular}
  \end{center}
  \vspace{-5mm}
\end{table}

\subsection{The Out-performance of HateModerate over Existing Datasets}
\label{sec:HM_vs_DH}

 To address RQ4, we need to demonstrate that HateModerate outperforms DynaHate. HateModerate is constructed by reusing existing examples from DynaHate and other datasets, dominated by DynaHate. We carry out a comparative analysis focusing on the failure rates of two pipelines: \textcircled{1}: Fine-tuning with CardiffNLP data + DynaHate. \textcircled{2}: Fine-tuning with CardiffNLP data + DynaHate + (HateModerate - DynaHate). Details of the experiment are provided in Appendix~\ref{sec:HM_vs_DH_appendix}. 
 The T-tests Table~\ref{tab:p-values} indeed show that Pipeline \textcircled{1} significantly outperforms Pipeline \textcircled{2} on both the HateCheck and the HateModerate test set:  



\begin{table}[ht]
\centering
\begin{tabular}{lccc}
\toprule
\textbf{Test Set} & \textbf{Category} & \textbf{p-value} \\
\midrule
HateCheck & Overall & $1.4701 \times 10^{-8}$ \\
HateCheck & Hate & $1.5947 \times 10^{-15}$ \\
HateCheck & Non-Hate & $0.0202$ \\
HateModerate & Overall & $0.0080$ \\
HateModerate & Hate & $2.7837 \times 10^{-8}$ \\
HateModerate & Non-Hate & $0.0016$ \\
\bottomrule
\end{tabular}
\caption{The p-values indicating the statistical significance of performance differences between pipeline \textcircled{1} and \textcircled{2}.}
\label{tab:p-values}
\end{table}

\vspace{-0.1in}

\section{Conclusions}
\label{sec:conclusion}

In this paper, we propose a dataset HateModerate, which includes hateful and non-hateful examples matching the 41 community standards guideline policies of Facebook. First, we leverage manual annotation with 28 graduate students followed by information retrieval, data augmentation, and verification to construct a dataset containing both hateful and non-hateful examples. Second, we use HateModerate to test the failures of state-of-the-art hate detection models. We find that popular content moderation models frequently make mistakes for both hateful and non-hateful examples. Finally, we observe that by augmenting the training data with HateModerate, the model can better conform to HateModerate while having the comparable performance to the original test data. Our study highlights the importance of investigating hate speech detectors' conformity to content policies.

\newpage

\section{Limitations}
\label{sec:limitation}

\noindent \textbf{Extending HateModerate to New Policies}. HateModerate is built based on Facebook's content moderation policy on Nov 23, 2022~\cite{fb_standard}. When applying our work to different policies (e.g., for a different platform), we must hire new human annotators to search for the matching examples. One future direction for improving this limitation is to automatically retrieve the matching examples given the policy. 


\noindent \textbf{Comprehensiveness of Content Policies}. Although Facebook's content moderation policies on hate speech are relatively comprehensive, the 41 policies may not cover all hate speech. 

\noindent \textbf{Mitigating the Data Bias of HateModerate}. Our data collection leverages searches based on community standards guidelines. Since the searches are initiated based on the guidelines, the collected dataset may contain bias in the following aspects. First, the data might be skewed towards keywords explicitly mentioned or can be easily inferred from the guideline. Second, the dataset may contain limited \emph{implicit hateful sentences}. One way to mitigate the first bias is to enumerate concepts given the high-level guideline, e.g., by querying the GPT engine: "\emph{Enumerate a list of objects (i.e., things) for the dehumanization of women:  }". For the second bias, following previous work on implicit hateful examples~\cite{elsherief2021latent}, we plan to explore automated categorization to improve the coverage of implicit hate in HateModerate.

\section{Ethics Considerations}
\label{sec:ethics}

\textbf{License/Copyright}. HateModerate primarily relies on reusing examples from existing hate speech data including DynaHate~\cite{vidgen2020learning} and HateCheck~\cite{rottger2020hatecheck}. We refer users to the original licenses accompanying each dataset.

\noindent \textbf{Intended Use}. HateModerate's intended use is as an evaluation tool for hate speech detection models, supporting capability tests to help diagnose model failures. We demonstrated this use of HateModerate in Section~\ref{sec:step2}. We also briefly discussed alternative uses of HateModerate in Section~\ref{sec:future}, e.g., as a dataset for explaining a decision for hate moderation by linking the decision to one of the content policies. These uses aim at aiding the development of better hate speech detection models. HateModerate reuses existing hate speech datasets including DynaHate~\cite{vidgen2020learning} and HateCheck~\cite{rottger2020hatecheck}, and our usage for these datasets is consistent with the intended use described in their papers.

\noindent \textbf{Potential Misuse}. Similar to existing datasets for capability tests~\cite{rottger2020hatecheck}, one potential misuse is over-extending claims about the functionalities of hate detection models. Our dataset may allow malicious actors to generative model that can generate hate speech matching the requirement for specific policies, which may further help them attack existing content moderators in a more structured manner. Nevertheless, due to the small scale of our dataset, this will unlikely happen. Overall, the scientific and social benefits of the research arguably outweigh the small risk of their misuse. 

\noindent \textbf{Annotator Recruitment and Compensation}.
\label{subsec:annotators}
HateModerate is annotated by 28 graduate students (10 Indian, 9 Chinese, 9 USA) in Computer Science, all of them are fluent English speakers. The student annotators in this paper are recruited from PhD and Master students at a research lab and students taking a graduate-level NLP course. They were rewarded \$20 Amazon e-gift cards as compensation for their annotation efforts. The entire annotation process spans seven months while the actual annotation time takes about seven weeks (four weeks for hate, three weeks for non-hate). The annotator names are anonymized in the dataset. We obtained the annotators' consent, and it was explained to the annotators how the data would be used.


\clearpage
\bibliography{anthology,custom}


\clearpage

\appendix
\section{Appendix}
\label{sec:appendix}

\subsection{Case Studies of Policies with High Failure Rates}
\label{sec:casestudy}

\begin{itemize}
\item Tier 1: "\emph{Dehumanizing speech as Certain Objectsects (women as household objects or property or objects in general; Black people as farm equipment; transgender or non-binary people as “it”)}": In particular, most models fails to detect hate speeches such as "\emph{MAKE WOMEN PROPERTY AGAIN!!!!}".

\item Tier 2: "\emph{Expressions of contempt, self-admission to intolerance on the basis of protected characteristics}": e.g., Perspective fail to detect "\emph{I am racist because I'm a white person grown up in white American systems.}".

\item Tier 3: "\emph{Economic exclusion, which means denying access to economic entitlements and limiting participation in the labour market}": e.g., "\emph{Pregnant women should not get any promotion at work as they will end up in maternity leave}".

\item Tier 4: "\emph{Content explicitly providing or offering to provide products or services that aim to change people’s sexual orientation or gender identity}" We find all models fail in the half of this policy, e.g., "\emph{With the right therapy and support gays can be cured and live much more fulfilling lives}".
\end{itemize}

\subsection{The Classes of Non-Hateful Examples}
\label{sec:class_nonhate}
\begin{itemize}
    \item Counter hate or referencing hate, e.g., "\emph{It is not correct to say 'Muslims should be killed'.}";
    \item Offensive towards non-protected entities, e.g., "\emph{I hate cauliflower.}";
    \item Neutral, e.g., "\emph{Black people have darker skins.}";
    \item Supportive, e.g., "\emph{Women should be promoted.}";
    \item Mislabeled examples, i.e., hateful or the example does not match the policy;
\end{itemize}

\subsection{The List of Evaluated Hate Speech Models}
\label{sec:models}

\begin{itemize}
\item Google's Perspective API: Jigsaw's Perspective API~\cite{perspective_api} is a popular model for content moderation, which is frequently used in downstream moderation tasks including news publishing, social media~\cite{perspective_api_use}, as well as helping social and political science research~\cite{friedl2023dis}. Perspective leverages training data from a variety of sources, including comments from online forums such as Wikipedia and The New York Times\footnote{\url{https://developers.perspectiveapi.com/s/about-the-api-training-data?language=en_US}}.

\item OpenAI's Moderation API: OpenAI's Moderation API~\cite{openai_moderation} OpenAI's content moderation endpoint, it is based on a GPT model fine-tuned using the classification head as the objective function~\cite{markov2022holistic}. The fine-tuning leverages both public hate speech datasets and the production data of OpenAI, and it requires continuous training to adapt to the new hateful content~\cite{markov2022holistic}. This model is being actively maintained and has been used by Stanford's Alpaca to improve the safety alignment of the text generation~\cite{alpaca}. 

\item Cardiff NLP's Fine-Tuned RoBERTa model: This open-source model is a fine-tuned RoBERTa model by Cardiff University's NLP group~\cite{antypas2023robust}. The complete list of the 13 datasets used for fine-tuning can be found on the model's HuggingFace page: \cite{cardiffnlp}. The older version of this model is the top-2 most downloaded fine-tuned model (84.6k downloads as of Oct 2023) for English hate-speech detection on the HuggingFace platform~\footnote{\url{https://huggingface.co/models?sort=downloads&search=hate}}. 

\item Facebook's Fine-Tuned RoBERTa model~\cite{fb_moderation_api_r1}: This open-source model is a fine-tuned RoBERTa model by Facebook and the Alan Turing Institute~\cite{fb_moderation_api_r1}. The fine-tuning leverages 11 datasets, although the exact list is not revealed by the authors~\cite{vidgen2020learning}. The R4 version of this model is the top-1 most downloaded fine-tuned model (54k downloads as of Oct 2023) for English hate-speech classification on HuggingFace. Instead of R4, we evaluate the R1 model, because the R4 model is fine-tuned on DynaHate thus evaluating R4 causes the data contamination problem~\cite{magar2022data}.

\end{itemize}

\subsection{The List of the 9 Training Datasets for CardiffNLP's Model}
\label{sec:card_datasets}

Although the CardiffNLP model uses 13 datasets for fine-tuning~\cite{antypas2023robust}, 4 datasets are non-downloadable, we list the 9 accessible datasets below: 

\begin{itemize}
\item \textbf{Measuring hate speech (MHS)}~\cite{sachdeva2022measuring} include 39,565 social media comments.
\item \textbf{Call me sexist, but (CMS)}~\cite{Samory2020CallMS} consist of 6,325 sentences related with sexism.
\item \textbf{Hate Towards the Political Opponent (HTPO)}~\cite{grimminger-klinger-2021-hate} collect 3,000 tweets about the 2020 USA president election.
\item \textbf{HateXplain}~\cite{mathew2021hatexplain} contains 20,148 posts from Twitter/X and Gab.
\item \textbf{Offense}~\cite{zampieri-etal-2019-predicting} is a collection of 14,100 tweets about offensive or non-offensive.
\item \textbf{Automated Hate Speech Detection (AHSD)}~\cite{davidson2017automated} combine 24,783 tweets.
\item \textbf{Multilingual and Multi-Aspect Hate Speech Analysis (MMHS)}~\cite{ousidhoum-etal-2019-multilingual} is a dataset with 5,647 tweets in three different languages: English, Arabic and French.
\item \textbf{HatE}~\cite{basile-etal-2019-semeval} is a collection of 19,600 tweets with English and Spanish languages.
\item \textbf{Detecting East Asian Prejudice on Social Media (DEAP)}~\cite{vidgen-etal-2020-detecting} has 20,000 tweets which focus on East Asian prejudice.
\end{itemize}

\subsection{Excluding Sentences to Prevent Data Contamination}
\label{sec:remove_original}

In this paper, to reduce the risk of data contamination, i.e., overlaps between the train and test dataset, we need to exclude the examples from HateModerate that can potentially exist in the training data of the evaluated models. First, OpenAI API and Google Perspective have not released their training sets. Second, among the training datasets of CardiffNLP~\cite{antypas2023robust}, we identify that Waseem et al.~\cite{waseem2016you} and Founta et al.~\cite{founta2018large} are used in DynaHate's R0 dataset~\cite{vidgen2020learning}. As a result, we exclude all examples in DynaHate that are originally from other datasets and only keep those that are newly created. More specifically, we keep only the perturbed examples in rounds 2, 3, and 4. Finally, since Facebook's training datasets have no overlaps with the DynaHate, there is little risk of data contamination with HateModerate.


\subsection{The Hypeparameters and Details of the Fine-Tuning Process}
\label{sec:finetune_process}

To study the effectiveness of HateModerate in reducing models' non-conformity issues, we fine-tune two RoBERTa models: \textcircled{1} Fine-tuning using the CardiffNLP 9 datasets in Section~\ref{sec:card_datasets}; \textcircled{2} Fine-tuning using CardiffNLP datasets + HateModerate. The hyperparameter tuning process explores a range of learning rates and epoch sizes. Specifically, we experiment with grid search using the learning rates ${1E-5, 2E-5}$, epoch sizes ${2, 3, 4}$, and training batch size ${4, 16, 32}$. For both models, the warm-up steps are $50$. The grid search space is chosen by referring to the best-performed hyperparameters setting of Cardiff NLP models as described in~\cite{antypas2023robust}. The best-identified hyperparameters for both models are learning rate $= 2E-5$, batch size $=32$, and epoch size $=4$. Both models are fine-tuned on a server with 4x NVIDIA V100 GPUs, the training takes approximately half an hour per epoch for both models.


\subsection{Comparison of Model Failures of Different Sub-Categories of Non-Hateful Speeches}
\label{sec:sub_nonhate}

\begin{figure}[h]
    \centering
    \includegraphics[width=0.95\linewidth]{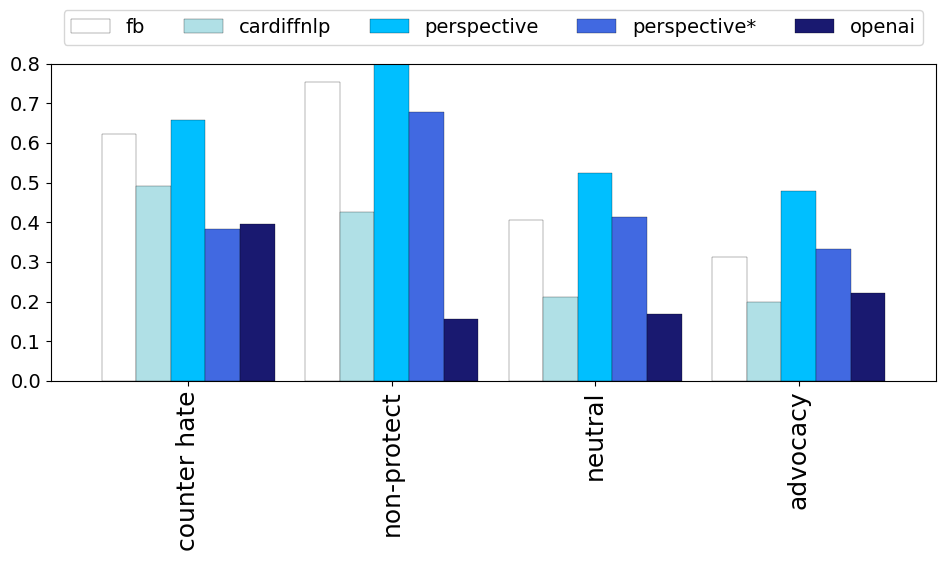}
    \caption{The comparison of failure rates in each sub-categories of non-hateful examples}
    \label{fig:counterhate}
\end{figure}

To better understand the failures in non-hateful examples, we further conduct a comparative study on the failure rates between different sub-categories of the non-hateful examples. We show the results in Figure~\ref{fig:counterhate}. Among all the 4 non-hateful categories, we find that counter hate and attacking non-protected group has the highest failure rate, whereas advocating for protected groups has the lowest failure rate. This result is consistent with our expectation, since the former categories sound more aggressive. 


 \begin{table}[h]
\caption{Measuring the bias: Simple test cases in which we find the hate probability of the fine-tuned model abnormal. The numbers in this table are the hate probabilities for each example. \label{table:simple_test} }
\begin{center}
\vspace{-5mm}
  \label{table:simple_test}
  \begin{tabular}{p{3cm}p{1cm}p{1cm}p{1cm}}
    \toprule
    Test case & \textbf{Cardiff} & \textbf{+HM} & \textbf{+HM$^*$} \\
    \midrule\\
    Middle Eastern people are selfish. &\textcolor{red}{\textbf{0.001}} & \textcolor{orange}{\textbf{0.748}} & \textcolor{red}{\textbf{0.414}} \\
    Asians are selfish. & \textcolor{orange}{\textbf{0.795}} & 0.997 &0.981 \\
    Homos are ugly. & 0.986 & \textcolor{red}{\textbf{0.218}} & \textcolor{orange}{\textbf{0.765}} \\
    \bottomrule
  \end{tabular}
  \end{center}
  \vspace{-5mm}
\end{table}

\subsection{Details on the Significance Tests for the Fine-Tuning Experiments}
\label{sec:ttest}

For the fine-tuning experiments in Table~\ref{tab:fine_tune_roberta}, we perform paired t-tests\footnote{\url{https://docs.scipy.org/doc/scipy/reference/generated/scipy.stats.ttest_rel.html}} between \textbf{CardiffNLP} vs \textbf{+HM} and \textbf{CardiffNLP} vs \textbf{+HM$^*$}. The statistics and p-values of the t-tests are shown in Table~\ref{tab:ttest}. For each t-test, if the statistics is positive, it means the CardiffNLP baseline performs better and vice versa. The results where \textbf{+HM} or \textbf{+HM$^*$} significantly outperforms \textbf{CardiffNLP} are denoted in bold.

\begin{table}[t]
\centering
\caption{The statistics and p-values of the paired t-tests for comparing fine tuning with and without HateModerate\label{tab:ttest}}
\small
\renewcommand{\arraystretch}{1.0}
\setlength\tabcolsep{3.5pt} 
\begin{tabular}{ccccc}

\toprule & \multicolumn{2}{l}{\textbf{Card} vs \textbf{+HM}} & \multicolumn{2}{l}{\textbf{Card} vs \textbf{+HM$^*$}}\\
& statistics & p-value & statistics &p-value\\
\midrule
\multicolumn{5}{l}{\textit{\textbf{HateCheck~\cite{rottger2020hatecheck}}}} \\
Hate & \textbf{25.59}& \textbf{1.1E-133} & \textbf{20.43}& \textbf{3.0E-88}\\
Non-hate & -2.51& 1.2E-02 & -0.43& 6.7E-1\\
Overall & \textbf{23.90}& \textbf{6.0E-118} & \textbf{18.09}& \textbf{2.7E-70}\\
\midrule
\multicolumn{5}{l}{\textit{\textbf{HateModerate Test}}} \\
Hate & \textbf{20.79}& \textbf{2.9E-91} & \textbf{15.66}& \textbf{1.0E-53}\\
Non-hate & \textbf{5.85}& \textbf{5.4E-09} & \textbf{7.82}& \textbf{6.7E-15}\\
Overall & \textbf{12.11}& \textbf{3.7E-33} & \textbf{6.47}& \textbf{1.1E-10}\\
\midrule
\multicolumn{5}{l}{\textbf{CardiffNLP Test Sets:}} \\
\multicolumn{5}{l}{\textit{\textbf{hatEval~\cite{basile-etal-2019-semeval}}}} \\

Hate & 1.18
& 2.4E-01 & \textbf{3.31}& \textbf{9.4E-04}\\
Non-hate & -0.61
& 5.4E-01& -0.44& 6.6E-01\\
Overall & 1.19& 2.4E-01 & \textbf{2.17}& \textbf{3.0E-02}\\
\multicolumn{5}{l}{\textit{\textbf{HTPO~\cite{grimminger-klinger-2021-hate}}}} \\
Hate & -2.32
& 2.1E-02
& 0.00
& 1.0E+00
\\
Non-hate & 0.73
& 4.7E-01
& 0.21
& 8.4E-01
\\
Overall & -2.05& 4.1E-02 & -0.16& 8.7E-01\\
\multicolumn{5}{l}{\textit{\textbf{HateXplain~\cite{mathew2021hatexplain}}}} \\
Hate & -0.34& 7.4E-01
& -1.31& 1.9E-01\\
Non-hate & \textbf{3.71}&\textbf{ 2.1E-04}
& \textbf{3.63}& \textbf{2.9E-04}
\\
Overall & -3.10& 1.9E-03& -3.54&  4.1E-04\\
\bottomrule
\end{tabular}
\end{table}

\clearpage
\onecolumn

\subsection{Comparative Analysis of HateModerate and Dynahate}
\label{sec:HM_vs_DH_appendix}

We conduct a comparative study that compares the failure rates of two pipelines: \textcircled{1}: Fine-tuning with CardiffNLP data + DynaHate. \textcircled{2}: Fine-tuning with CardiffNLP data + DynaHate + (HateModerate - DynaHate).
In Table~\ref{tab:HM_vs_DH}, we can observe that fine-tuning with HateModerate outperforms fine-tuning with DynaHate; the improvement is large on the first few guidelines.

\begin{longtable}{|p{1.5cm}|p{5.5cm}|p{1.7cm}|p{1.7cm}|p{1.7cm}|p{1.7cm}|}
\captionsetup{justification=raggedright, singlelinecheck=false, width=\textwidth}
\caption{Failure rates of \textcircled{1} and \textcircled{2} on all categories of HateModerate. \textcircled{1}: Fine-tuning with CardiffNLP data + DynaHate. \textcircled{2}: Fine-tuning with CardiffNLP data + DynaHate + (HateModerate - DynaHate)}
\label{tab:HM_vs_DH}
\\
\hline
\textbf{\# cases in DH} & \textbf{Guideline} & \textbf{Method \textcircled{1} (hate)} & \textbf{Method \textcircled{2} (hate)} & \textbf{Method \textcircled{1} (all)} & \textbf{Method \textcircled{2} (all)} \\
\hline

All & All & 0.210 & \textbf{0.097} & 0.191 & \textbf{0.102} \\
\hline
3 & 38 - change sexual & 0.735 & \textbf{0.245} & 0.379 & \textbf{0.198} \\
\hline
7 & 36 - economic exclusion & 0.466 & \textbf{0.023} & 0.326 & \textbf{0.056} \\
\hline
 8 & 24 - contempt self admission intolerance & 0.539 & \textbf{0.022} & 0.405 & \textbf{0.074} \\
\hline
17 & 10 - certain objects & 0.500 & \textbf{0.000} & 0.259 & \textbf{0.155} \\
\hline
24 & 4 - disease & 0.111 & \textbf{0.000} & 0.089 & \textbf{0.067} \\
\hline
29 & 35 - political exclusion & 0.481 & \textbf{0.148} & 0.367 & \textbf{0.200} \\
\hline
32 & 11 - deny existence & 0.167 & \textbf{0.071} & 0.188 & \textbf{0.088} \\
\hline
39 & 25 - contempt shouldn't exist & 0.190 & \textbf{0.095} & 0.109 & \textbf{0.065} \\
\hline
52 & 12 - harmful stereotype & 0.182 & \textbf{0.091} & 0.162 & \textbf{0.081} \\
\hline
57 & 18 - attack mental health & 0.160 & \textbf{0.000} & 0.162 & \textbf{0.027} \\
\hline
58 & 7 - sexual predator & \textbf{0.081} & 0.108 & \textbf{0.146} & 0.167 \\
\hline
65 & 17 - attacking education & 0.143 & \textbf{0.057} & 0.140 & \textbf{0.070} \\
\hline
67 & 3 - bacteria & 0.455 & \textbf{0.091} & 0.294 & \textbf{0.059} \\
\hline
72 & 0 - filth & 0.053 & \textbf{0.026} & 0.043 & \textbf{0.022} \\
\hline
75 & 19 - attacking character trait & 0.086 & \textbf{0.000} & 0.085 & \textbf{0.000} \\
\hline
77 & 14 - attacking hygiene & \textbf{0.000} & 0.037 & \textbf{0.000} & 0.026 \\
\hline
77 & 29 - disgust vomit & 0.308 & \textbf{0.269} & 0.290 & \textbf{0.226} \\
\hline
77 & 37 - social exclusion & 0.182 & \textbf{0.182} & 0.200 & \textbf{0.150} \\
\hline
78 & 26 - contempt despise hate & 0.333 & \textbf{0.222} & 0.368 & \textbf{0.263} \\
\hline
88 & 31 - curse genitalia & 0.095 & \textbf{0.071} & 0.104 & \textbf{0.063} \\
\hline
89 & 33 - segregation & 0.231 & \textbf{0.179} & 0.200 & \textbf{0.150} \\
\hline
104 & 27 - contempt despise dislike & 0.318 & \textbf{0.273} & 0.290 & \textbf{0.194} \\
\hline
106 & 5 - dehumanization animal & 0.026 & \textbf{0.000} & 0.064 & \textbf{0.000} \\
\hline
109 & 15 - attacking appearance & 0.026 & \textbf{0.026} & 0.043 & \textbf{0.021} \\
\hline
112 & 6 - feces & 0.121 & \textbf{0.052} & 0.113 & \textbf{0.048} \\
\hline
123 & 30 - disgust repulsion & 0.158 & \textbf{0.105} & 0.159 & \textbf{0.114} \\
\hline
123 & 2 - insects & 0.484 & \textbf{0.422} & 0.421 & \textbf{0.355} \\
\hline
129 & 8 - subhumanity & 0.259 & \textbf{0.185} & 0.222 & \textbf{0.167} \\
\hline
134 & 9 - criminal & 0.297 & \textbf{0.270} & 0.244 & \textbf{0.244} \\
\hline
135 & 39 - attack concept associated protected characteristics & 0.353 & \textbf{0.176} & 0.207 & \textbf{0.138} \\
\hline
135 & 28 - curse sexual & 0.079 & \textbf{0.048} & 0.111 & \textbf{0.069} \\
\hline
135 & 34 - explicit exclusion & 0.086 & \textbf{0.034} & 0.113 & \textbf{0.113} \\
\hline
135 & 21 - less than adequate & 0.064 & \textbf{0.043} & 0.055 & \textbf{0.036} \\
\hline
137 & 23 - better worse than & 0.327 & \textbf{0.041} & 0.269 & \textbf{0.060} \\
\hline
141 & 16 - attacking intellectual capability & 0.119 & \textbf{0.119} & 0.113 & \textbf{0.113} \\
\hline
142 & 20 - attacking derogatory term & 0.093 & \textbf{0.037} & 0.077 & \textbf{0.046} \\
\hline
148 & 32 - curse profane & \textbf{0.029} & 0.043 & \textbf{0.027} & 0.040 \\
\hline
149 & 22 - deviating norm & 0.145 & \textbf{0.127} & 0.111 & \textbf{0.097} \\
\hline
151 & 40 - spread virus & 0.241 & \textbf{0.193} & 0.250 & \textbf{0.198} \\
\hline
189 & 1 - violent speech & 0.048 & \textbf{0.016} & 0.055 & \textbf{0.014} \\
\hline
281 & 13 - hatecrime victim & 0.105 & \textbf{0.060} & 0.103 & \textbf{0.062} \\
\hline
\end{longtable}



\subsection{Overview of Facebook's Hate Speech Community Standards}
\label{sec:fb_standards}

In Table~\ref{tab:fb_standards}, we provide a copy of Facebook's community standards guidelines as of Nov 2022. The guidelines also serve as the text instructions provided to the human annotators for all six steps of data labeling in Section~\ref{sec:step1} (Figure~\ref{fig:workflow}).

\begin{longtable}{|p{0.4cm}|p{0.5cm}|p{2cm}|p{12cm}|}
\captionsetup{justification=raggedright, singlelinecheck=false, width=\textwidth}
\caption{Full name and description for Facebook's Hate Speech Community Standards~\cite{fb_standard}. We show matching names of guidelines and their index in Figure~\ref{fig:fail_rate_all}, the full descriptions of them are following.\label{tab:fb_standards} } \\
\hline
\textbf{ID} & \textbf{Tier} & \textbf{Guideline} & \textbf{Description} \\
\hline
0 & 1 & \textsf{Dehumanize Filth} & Dehumanizing speech: Filth (including but not limited to: dirt, grime) \\ \hline
1 & 1 & \textsf{Violent Speech} & Violent speech or support in written or visual form \\ \hline
2 & 1 & \textsf{Dehumanize Insects} & Dehumanizing speech: Insects (including but not limited to: cockroaches, locusts) \\ \hline
3 & 1 & \textsf{Dehumanize Bacteria} & Dehumanizing speech: Bacteria, viruses, or microbes \\ \hline
4 & 1 & \textsf{Disease} & Dehumanizing speech: Disease (including but not limited to: cancer, sexually transmitted diseases) \\ \hline
5 & 1 & \textsf{Dehumanize Animals} & Dehumanizing speech: Animals in general or specific types of animals that are culturally perceived as intellectually or physically inferior (including but not limited to: Black people and apes or ape-like \\ \hline
6 & 1 & \textsf{Feces} & Dehumanizing speech: Feces (including but not limited to: shit, crap) \\ \hline
7 & 1 & \textsf{Sexual Predator} & Dehumanizing speech: Sexual predators (including but not limited to: Muslim people having sex with goats or pigs) \\ \hline
8 & 1 & \textsf{Subhumanity} & Dehumanizing speech: Subhumanity (including but not limited to: savages, devils, monsters, primitives) \\ \hline
9 & 1 & \textsf{Criminal} & Violent criminals (including but not limited to: terrorists, murderers, members of hate or criminal organizations). Other criminals (including but not limited to “thieves,” “bank robbers,” or saying “All [protected characteristic or quasi-protected characteristic] are ‘criminals’”). \\ \hline
10 & 1 & \textsf{Certain Objects} & Certain objects (women as household objects or property or objects in general; Black people as farm equipment; transgender or non-binary people as “it”) \\ \hline
11 & 1 & \textsf{Deny Existence} & Statements denying existence (including but not limited to: "[protected characteristic(s) or quasi-protected characteristic] do not exist", "no such thing as [protected charactic(s) or quasi-protected characteristic]" ), deny existence is different from contempt-should-not-exist in tier 2 \\ \hline
12 & 1 & \textsf{Harmful Stereotype} &  Harmful stereotypes historically linked to intimidation, exclusion, or violence on the basis of a protected characteristic, such as Blackface; Holocaust denial; claims that Jewish people control financial, political, or media institutions; and references to Dalits as menial laborers\\ \hline
13 & 1 & \textsf{Hatecrime Victim} & Mocking the concept, events or victims of hate crimes even if no real person is depicted in an image. \\ \hline
14 & 2 & \textsf{Attack Hygiene} & Content attacking a person or a group based on their hygiene, including but not limited to: filthy, dirty, smelly. \\ \hline
15 & 2 & \textsf{Attack Appearance} & Content attacking a person or a group based on their physical appearance, including but not limited to: ugly, hideous. \\ \hline
16 & 2 & \textsf{Attack Intellectual Capability} & Content attacking a person or a group based on their intellectual capacity, including but not limited to: dumb, stupid, idiots. \\ \hline
17 & 2 & \textsf{Attack Education} & Content attacking a person or a group based on their education, including but not limited to: illiterate, and uneducated. \\ \hline
18 & 2 & \textsf{Attack Mental Health} & Content attacking a person or a group based on their mental health, including but not limited to: mentally ill, retarded, crazy, insane. \\ \hline
19 & 2 & \textsf{Attack Characteristics} & Attacking a person or group based on a perceived character trait, Content attacking a person or a group based on their character traits culturally perceived as negative, including but not limited to: coward, liar, arrogant, ignorant. \\ \hline
20 & 2 & \textsf{Attack Derogatory} & Content attacking a person or a group based on derogatory terms related to sexual activity, including but not limited to: whore, slut, and perverts. \\ \hline
21 & 2 & \textsf{Less Than Adequate} & Expressions about being less than adequate, including but not limited to: worthless, useless \\ \hline
22 & 2 & \textsf{Deviating Norm} & Expressions about deviating from the norm, including but not limited to: freaks, abnormal. \\ \hline
23 & 2 & \textsf{Better Worse Than} & Expressions about being better/worse than another protected characteristic, including but not limited to: "I believe that males are superior to females." \\ \hline
24 & 2 & \textsf{Contempt Self Admission Intolerance} & Self-admission to intolerance on the basis of a protected characteristic, including but not limited to: homophobic, islamophobic, and racist. \\ \hline
25 & 2 & \textsf{Contempt Shouldn't Exist} & Expressions that a protected characteristic shouldn't exist. (shouldn't exist is different from deny-existence in tier 1) \\ \hline
26 & 2 & \textsf{Contempt Despise Hate} & Expressions of hate, including but not limited to: despise, hate. \\ \hline
27 & 2 & \textsf{Contempt Despise Dislike} & Expressions of dismissal, including but not limited to: don´t respect, don't like, don´t care for \\ \hline
28 & 2 & \textsf{Curse Sexual} & Terms or phrases calling for engagement in sexual activity, or contact with genitalia, anus, Feces or urine, including but not limited to: suck my dick, kiss my ass, eat shit. \\ \hline
29 & 2 & \textsf{Disgust Vomit} & Expressions that suggest the target causes sickness, including but not limited to: vomit, throw up. \\ \hline
30 & 2 & \textsf{Disgust Repulsive} & Expressions of repulsion or distaste, including but not limited to: vile, disgusting, yuck. \\ \hline
31 & 2 & \textsf{Curse Genitalia} & Curse that referring to the target as genitalia or anus, including but not limited to: cunt, dick, asshole. \\ \hline
32 & 2 & \textsf{Curse Profane} & Profane terms or phrases with the intent to insult, including but not limited to: fuck, bitch, motherfucker. \\ \hline
33 & 3 & \textsf{Segregation} & Segregation in the form of calls for action, statements of intent, aspirational or conditional statements, or statements advocating or supporting segregation. \\ \hline
34 & 3 & \textsf{Explicit Exclusion} & Call for action of exclusion, e.g., explicit exclusion, which means things like expelling certain groups or saying they are not allowed. \\ \hline
35 & 3 & \textsf{Political Exclusion} & Call for action of exclusion, e.g., political exclusion, which means denying the right to political participation. \\ \hline
36 & 3 &\textsf{Economic Exclusion} & Call for action of exclusion, e.g., economic exclusion, which means denying access to economic entitlements and limiting participation in the labour market. \\ \hline
37 & 3 & \textsf{Social Exclusion} &  Call for action of exclusion, e.g., social exclusion, which means things like denying access to spaces (physical and online)and social services, except for gender-based exclusion in health and positive support Groups.\\ \hline
38 & 4 & \textsf{Change Sexual} & Content explicitly providing or offering to provide products or services that aim to change people’s sexual orientation or gender identity. \\ \hline
39 & 4 & \textsf{Attack Concepts} &  Content attacking concepts, institutions, ideas, practices, or beliefs associated with protected characteristics, which are likely to contribute to imminent physical harm, intimidation or discrimination against the people associated with that protected characteristic.\\ \hline
40 & 4 & \textsf{Spread Virus} & Content targeting a person or group of people on the basis of their protected characteristic(s) with claims that they have or spread the novel coronavirus, are responsible for the existence of the novel coronavirus, are deliberately spreading the novel coronavirus, or mocking them for having or experiencing the novel coronavirus. \\
\hline
\end{longtable}

\end{document}